\documentclass[sigplan,10pt,screen]{acmart}

\usepackage[utf8]{inputenc}
\usepackage{xspace}
\usepackage{enumitem}
\setlist[itemize]{noitemsep}
\usepackage{tikz}
\usetikzlibrary{fit}
\usepackage{listings}
\usepackage{mathpartir}
\usepackage{amsmath}

\usepackage{amssymb}

\definecolor{egreen}{rgb}{0, 0.4, 0.267}
\definecolor{dkviolet}{rgb}{0.6,0,0.8}
\definecolor{dkgreen}{rgb}{0,0.4,0}
\definecolor{dkblue}{rgb}{0,0.1,0.5}
\definecolor{lightblue}{rgb}{0,0.5,0.5}
\definecolor{orange}{rgb}{0.9,0.39,0}

\newcommand{\HL}[1]{\textcolor{orange}{#1}}
\newcommand{\HLb}[1]{\textcolor{dkblue}{#1}}

\newcommand{\code}[1]{\lstinline!#1!}
\newcommand{\elpicode}[1]{\lstinline[language=ELPI]!#1!}

\newcommand{\Sec}{\S} 

\newcommand{\ie}{\textit{i.e.}, }
\newcommand{\eg}{\textit{e.g.}, }

\newcommand{\namefont}[1]{\texttt{#1}}

\newcommand{\MetaCoq}{\namefont{MetaCoq}\xspace}
\newcommand{\SMTCoq}{\namefont{SMTCoq}\xspace}

\newcommand{\Ltac}{\namefont{Ltac}\xspace}
\newcommand{\CompCert}{\namefont{CompCert}\xspace}
\newcommand{\veriT}{\namefont{veriT}\xspace}

\newcommand{\cvcfive}{\namefont{CVC5}\xspace}
\newcommand{\Coq}{\namefont{Coq}\xspace}
\newcommand{\Agda}{\namefont{Agda}\xspace}
\newcommand{\Lean}{\namefont{Lean}\xspace}

\newcommand{\Prolog}{\namefont{Prolog}\xspace}
\newcommand{\lambdaProlog}{\namefont{$\lambda$Prolog}\xspace}
\newcommand{\Trakt}{\namefont{Trakt}\xspace}

\newcommand{\scope}{\texttt{scope}\xspace}
\newcommand{\Sniper}{\texttt{Sniper}\xspace}

\newcommand{\CoqHammer}{\texttt{CoqHammer}\xspace}
\newcommand{\fstar}{\texttt{F*}\xspace}

\newcommand{\coqelpi}{\texttt{Coq-Elpi}\xspace}
\newcommand{\elpi}{\texttt{Elpi}\xspace}
\newcommand{\MathComp}{\texttt{MathComp}\xspace}
\newcommand{\itauto}{\texttt{itauto}\xspace}

\newcommand{\snipe}{\lstinline+snipe+\xspace}
\newcommand{\verit}{\lstinline+verit+\xspace}

\newcommand{\firstordercongruence}{\lstinline+firstorder congruence+\xspace}
\newcommand{\trakt}{\lstinline+trakt+\xspace}

\newcommand{\Prop}{\lstinline+Prop+\xspace}
\newcommand{\bool}{\lstinline+bool+\xspace}

\newcommand{\LL}[1]{\mbox{\lstinline`#1`}}

\setcopyright{cc}
\acmPrice{15.00}
\acmDOI{10.1145/3573105.3575676}
\acmYear{2023}
\copyrightyear{2023}
\acmSubmissionID{poplws23cppmain-p24-p}
\acmISBN{979-8-4007-0026-2/23/01}
\acmConference[CPP '23]{Proceedings of the 12th ACM SIGPLAN International Conference on Certified Programs and Proofs}{January 16--17, 2023}{Boston, MA, USA}
\acmBooktitle{Proceedings of the 12th ACM SIGPLAN International Conference on Certified Programs and Proofs (CPP '23), January 16--17, 2023, Boston, MA, USA}
\received{2022-09-21}
\received[accepted]{2022-11-21}

\begin{document}

\title{Compositional pre-processing for automated reasoning in dependent type
  theory}

\author{Valentin Blot}
\affiliation{
  \institution{LMF, Inria, Université Paris-Saclay}
  \country{}
}

\author{Denis Cousineau}
\orcid{0000-0003-4078-3591}
\affiliation{
  \institution{Mitsubishi Electric R\&D Centre Europe}
  \country{}
}

\author{Enzo Crance}
\orcid{0000-0002-0498-0910}
\affiliation{
  \institution{LS2N, Inria, Nantes Université}
  \country{}
}
\affiliation{
  \institution{Mitsubishi Electric R\&D Centre Europe}
  \country{}
}

\author{Louise Dubois de Prisque}
\orcid{0000-0002-0389-0390}
\affiliation{
  \institution{LMF, Inria, Université Paris-Saclay}
  \country{}
}

\author{Chantal Keller}
\orcid{0000-0002-1282-0677}
\affiliation{
  \institution{LMF, Université Paris-Saclay}
  \country{}
}

\author{Assia Mahboubi}
\orcid{0000-0002-0312-5461}
\affiliation{
  \institution{LS2N, Inria, Nantes Université}
  \country{}
}

\author{Pierre Vial}
\affiliation{
  \institution{LMF, Inria, Université Paris-Saclay}
  \country{}
}

\renewcommand{\shortauthors}{V. Blot, D. Cousineau, E. Crance, L. Dubois
  de Prisque, C. Keller, A. Mahboubi, P. Vial}

\begin{abstract}
  In the context of interactive theorem provers based on a dependent type
theory, automation tactics (dedicated decision procedures, call of
automated solvers, ...) are often limited to goals which are exactly in
some expected logical fragment. This very often prevents users from applying
these tactics in other contexts, even similar ones.

This paper discusses the design and the implementation of
pre-processing operations for automating formal proofs in the \Coq
proof assistant. It presents the implementation of a wide variety of
predictible, atomic goal transformations, which can be composed in
various ways to target different backends. A gallery of examples
illustrates how it helps to expand significantly the power of
automation engines.

\end{abstract}

\ccsdesc[500]{Theory of computation~Automated reasoning}
\ccsdesc[500]{Theory of computation~Higher order logic}
\ccsdesc[500]{Theory of computation~Type theory}

\keywords{
interactive theorem proving,
Calculus of Inductive Constructions,
Coq,
automated reasoning,
arithmetic,
pre-processing}

\maketitle

\section{Introduction}
\label{sec:intro}

Mundane parts of formal proofs are best automated. But for users of the \Coq interactive theorem prover~\cite{the_coq_development_team_2022_5846982},
finding the appropriate plugin for doing so is not as simple as one
may wish for.  First, they have to find their way in the jungle of
available tools, with usually not much guarantee that it will be
robust to future irrelevant changes in the libraries. Moreover, some
slight changes in a statement, seemingly innocuous for a human, may
cause such a tool to fail. Generally speaking, it is hard to predict
failures or efficiency of a given automated reasoning proof command.

Actually, various powerful automated reasoning techniques have been
made available to interactive theorem proving. But once the user has
done all the high level reasoning (which can hardly be automated) and is
left with a goal that should be handled automatically, the bottleneck
does not lie so much in the power of automated reasoning
\textit{per se}, but rather concerns the pre-processing phase that
aligns the formula to be proved with the scope of the core automation
engine. The latter is indeed often tool-specific, hardly customizable
and little compositional. The lack of off-the-shelf goal transformations
for addressing this prerequisite actually significantly hampers proof
automation. The present paper discusses how to better design such
pre-processing phases, for improved formal proof automation.

The core idea is to put a suite of
\emph {independent}, \emph {atomic} goal transformations at the
service of an automation engine. In this model, a suitable
orchestration of small-scale transformations of various natures
(inductive reasoning, theory-specific translations, etc.) shall expand
the skills of the automation backend, while retaining predictability
and robustness of the latter. This model applies to a variety of
backend automation engines, and to an extensible collection of
small-scale transformations.

\paragraph{Why pre-processing?}
Proof commands, also called tactics, implementing formal proof
producing automated reasoning essentially fall in two camps, which
both involve preparing user goals before resorting to a core
automation engine. The first one consists of formal-proof-producing
implementations of decision procedures, \eg for the equational
theory of commutative rings~\cite{DBLP:conf/tphol/GregoireM05}, for
the first-order theory of real
numbers~\cite{DBLP:conf/cade/McLaughlinH05}, of
integers~\cite{DBLP:conf/types/Besson06} etc. These standalone formal
proof automation tactics typically target implementation-specific
choices of data structures or definitions, that are not be directly
compatible with users' choices. Goals involving data structures from
external libraries have thus to undergo a preliminary translation
phase before being passed to the core tactic~\cite{DBLP:conf/itp/Sakaguchi22}.

The second family targets a fragment of first-order logic, and is
often based on the integration of external automated theorem provers,
such as first-order provers, satisfiability (SAT) provers, or
satisfiability modulo theories (SMT) provers. In this case, the
untrusted output of these external provers is used to guide the
(re)construction of a formal proof. For instance,
\emph{hammers}~\cite{DBLP:conf/itp/DesharnaisVBW22}, as well as
\SMTCoq~\cite{DBLP:conf/cav/EkiciMTKKRB17} fall in this category.  But
interactive provers and automated provers usually do not speak the
same logic: most interactive theorem provers implement a flavor of
higher-order logic with inductive types, a strict superset of the
fragments of first-order logic handled by automated theorem
provers. As a consequence, hammers typically include a translation
heuristic from the logic of the interactive prover to that of the
automated prover, coupled with a formal proof reconstruction
mechanism. For instance, a general encoding of the Calculus of
Inductive Constructions into a target dialect of first order
logic~\cite{czajka:LIPIcs:2018:9853} grounds the implementation of a
hammer for the \Coq proof assistant.

 \paragraph{Structure.} The paper is organized as follows. Section~\ref{sec:examples}
  motivates this work by illustrating the zoo of tactics available for
  automated reasoning in the \Coq prover, and their
  limits. Section~\ref{sec:preproc} presents our main contribution: a
  collection of atomic,
  small-scale transformations for pre-processing goals. For instance,
  these transformations can explicity axiomatize inductive data types
  or ease theory-specific automation. Section~\ref{sec:applications}
  validates how such small-scale transformations, or a combination
  thereof, do enhance various automation backends.
  Section~\ref{sec:conclusion} provides some concluding remarks.

  The source code of the examples and the \Sniper plugin can be found
  at~\url{https://github.com/smtcoq/sniper/releases/tag/cpp23}. In
  particular, it contains a file \lstinline!examples/paper_examples.v!
  which presents all the examples of this paper, in the same order; it
  is designed to be executed throughout the reading of the paper. A
  \lstinline!README.md! explains how to build and execute the code. The
  source code of the \Trakt plugin can be found
  at~\url{https://github.com/ecranceMERCE/trakt/releases/tag/1.2%2B8.13}.

\section{Context and motivating examples}
\label{sec:examples}

Technical yet uninteresting proof steps are the daily
bread of program verification. Fortunately, many elementary
statements like the following fact:
\begin{lstlisting}
Lemma |*length_rev_app*| : forall B (l l' : list B),
  length (rev (l ++ l')) = length l + length l'.
\end{lstlisting}
about the length of a reversed appending of two lists are easily
solved by modern automated provers. The corresponding \emph{formal}
proof steps can in turn be automated, \eg using \emph{hammers}, a
powerful architecture for connecting external automated theorem
provers with formal interactive proof environments. The
\CoqHammer~\cite{DBLP:journals/jar/CzajkaK18} plugin equips \Coq{}
with an instance of hammer, inspired by the Isabelle/HOL pioneering
instance, but adapted to CIC. This plugin provides a \code{hammer}
tactic, which combines heuristics with calls to external provers for
first-order logic, so as to obtain hopefully sufficient hints,
including relevant lemmas from the current context, for proving the
goal. The actual formal proof is then reconstructed from these hints thanks 
to variants of the \code{sauto} tactic
and \code{hammer} outputs a corresponding robust and
oracle-independent proof script. For instance, it can produce the
script:
\begin{lstlisting}
Proof. scongruence use: app_length, rev_length. Qed.
\end{lstlisting}
for proving the \code{|*length_rev_app*|} example, using the
\code{app_length} and \code{rev_length} auxiliary lemmas from
the loaded standard library about lists. Yet, as of version 1.3.2, \CoqHammer is not
designed to exploit any theory-specific reasoning, and thus cannot
prove this slight variant, where \code{(b :: l')} replaces \code{l'}:
\begin{lstlisting}
Lemma |*length_rev_app_cons*| :
  forall B (l l' : list B) (b : B),
    length (rev (l ++ (b::l'))) =
    (length l) + (length l') + 1.
\end{lstlisting}
because it lacks arithmetical features. In this case, users may resort
to the \SMTCoq plugin~\cite{DBLP:conf/cav/EkiciMTKKRB17}, which
implements a certificate checker for proof witnesses output by
SMT solvers. The latter automated
provers are indeed tailored for finding proofs combining propositional
reasoning, congruence and theory-specific decision procedures, \eg
for linear arithmetic.  However, none of \CoqHammer or \SMTCoq can in
general reason by case analysis or induction. A variant of the
\code{sauto} tactic can prove the following fact about appending
though:
\begin{lstlisting}
Lemma |*app_nilI*| : forall B (l l' : list B),
  l ++ l' = [] -> l = [] /\ l' = [].
\end{lstlisting}
But neither \CoqHammer nor \SMTCoq can prove the following variant,
where the first list is reversed:
\begin{lstlisting}
Lemma |*app_nil_rev*| : forall B (l l' : list B),
  (rev l) ++ l' = [] -> l = [] /\ l' = [].
\end{lstlisting}
The \SMTCoq plugin can be used to prove properties of linear integer
arithmetic, but only when they are stated using the type \code{Z} of
integers from \Coq's standard library:
\begin{lstlisting}
Lemma |*eZ*| : forall (z : Z), z >= 0 -> z < 1 -> z = 0.
\end{lstlisting}
Up to version 2.0, \SMTCoq is however clueless about any
alternative instance of integer arithmetic, \eg the type \code{int}
of unary integers included in the Mathematical Components (or \MathComp)
library~\cite{assia_mahboubi_2021_4457887}:
\begin{lstlisting}
Lemma |*eint*| : forall (z : int), z >= 0 -> z < 1 -> z = 0.
\end{lstlisting}
Fortunately, \Coq distributes the \code{lia}
tactic~\cite{DBLP:conf/types/Besson06}, specialized to linear integer
arithmetic, which can actually also prove lemmas such as
\code{eZ}. Moreover, \code{lia} can be customized to a user-defined
instance of arithmetic thanks to the \code{zify} dedicated
pre-processing~\cite{ppsimpl}. Once correctly configured~\cite{mczify}
for type~\code{int}, \code{lia} is equally powerful on type \code{Z}
or type \code{int} and proves both \code{eZ} and \code{eint}. However, as
powerful as it may be on integer linear arithmetic, the \code{lia}
tactic is by nature unaware of the theory of equality. Hence, although
it can prove equality \code{eintC}, it is unable to prove the variant
\code{cong_eintC}, because the latter involves a congruence with the
\code{(_ :: nil)} operation, alien to the theory of linear integer
arithmetic:
\begin{lstlisting}
Lemma |*eintC*| : forall (z : int), z + 1 = 1 + z.
Lemma |*congr_eintC*| : forall (z : int),
  (z + 1) :: nil = (1 + z) :: nil.
\end{lstlisting}
Proving the property expressed by \code{congr_eintC} requires
\emph{combining} different theories, in this case integer arithmetic
and the theory of equality, as SMT solvers do. Yet, in this case as
well, the \SMTCoq plugin cannot help, because the statement of
this fact is phrased using type \code{int} instead of \code{Z}. The
recent \itauto SAT solver~\cite{DBLP:conf/itp/Besson21},
implemented in \Coq, provides an alternate take on formally verified
satisfiability modulo theory, and organizes the cooperation between
the independent tactics \code{lia}, for integer arithmetic, and
\code{congruence}, for equality. As a consequence, the \code{smt}
tactic built on top of \itauto can benefit from \code{lia}'s
pre-processing facilities. For instance, as soon as  \code{lia}'s
pre-processing is correctly configured for type \code{int},
the \code{smt} tactic is able to prove lemma
\code{congr_eintC}. However,
\code{lia}'s pre-processing facilities are not known to the rest of
the SMT decision procedure. Thus, although the
following goal is solved by the latter \code{smt} tactic:
\begin{lstlisting}
Lemma |*eintCb*| : forall (z : int), (z + 1 == 1 + z) = true.
\end{lstlisting}
because \code{lia} has been informed of the Boolean equality test
\code{(_ == _)} available on type \code{int}, the same tactic
fails on the \code{congr_eintCb} variant, featuring an uninterpreted symbol:
\begin{lstlisting}
Lemma |*congr_eintCb*| : forall (f : int -> int) (z : int),
  (f (z + 1) == f (1 + z)) = true.
\end{lstlisting}
As it turns out, although a variety of tactics implementing automated
reasoning is available to the users of the \Coq proof assistant,
finding the appropriate weapon for attacking a given goal remains challenging.
It is often quite difficult to anticipate the exact competence of
tactics based on first-order automated reasoning, and to interpret
failure. As a consequence, large-scale formalization endeavors may end
up developing their own specific automation tools, like the
\code{list_solve} tactic in the Verified Software
Toolchain~\cite{DBLP:conf/esop/Appel11}\cite[Chapters 64, 65]{VSTman},
for automating reasoning about lists and arithmetic, which makes
the number of available tactics multiply even more, often redundantly.

\section{Pre-Processing components}
\label{sec:preproc}

\subsection{The role of small-scale transformations}
\label{sec:preproc:transfos}
The pitfalls
illustrated in \Sec~\ref{sec:examples} pertain, generally, to the
distance between the standards of automated reasoning and the practice
of interactive theorem proving. These pitfalls are probably even more
acute in provers based on dependent type theory, and particularly so
in \Coq{}. For instance, in a prover like \Coq{}, or \Agda{}, or
\Lean{}, decidable theories may live in an ambient constructive logic,
and excluded middle for the corresponding formulas then follows from
this decidability result, rather than from a global axiom. Moreover,
\Coq{} in particular has a decentralized ecosystem of external
libraries, taking benefit from the versatility of the specification
language, and thus featuring different data structures, automation
tools, etc. Importantly, modern \Coq libraries often make use of
\emph{typeclass
mechanisms}~\cite{DBLP:conf/tphol/SozeauO08,DBLP:conf/itp/MahboubiT13}
for defining mathematical notations, which creates formulas with even
more syntactically different, but convertible, variants of a same
constant, \eg the addition on integers. And all these variants
should nonetheless be understood as the same symbol for automation to
succeed.

As these issues apply to any general-purpose automation plugin, the
corresponding solutions should remain as \emph{independent} as possible
from the targeted backend. What we refer to as \emph{pre-processing} is
the process which transforms a \Coq goal $C \vdash G$, with its local
context $C$, into a new goal $C' \vdash G'$ easier to handle for an
automation backend\footnote{We omit here the global context, as it is
  not affected by pre-processing.}. We expect it to be
\emph{proof-producing}, \ie to produce not only $C' \vdash G'$ but also
a Coq proof that the new goal entails the former one, to be checked by
Coq's kernel.

Pre-processing should be \emph{predictable}, in the sense that it should
not hamper the interpretation of failures (raised by the backend). A way
to preserve predictability is to avoid a silent use of the global
context, and sophisticated, heuristic-based proof-search. Pre-processing
transformations should also be well-specified, so as to avoid
unjustified discrepancies in the behavior of automation, which may
resemble unpredictability from the user's standpoint (\eg
\code{app_nilI} vs \code{app_nil_rev}, or \code{eintCb} vs
\code{congr_eintCb}). These precise specifications are a consequence of
the \emph{small-scale} and compositional nature of our
transformations\footnote{The compositionality of our transformations
  also makes the development process incremental and more robust.}.

Proof-producing, small-scale pre-processing transformations mostly come in two
flavors. The first one works by \emph{enriching the local context} $C$
of the initial goal into a larger $C'$, so as to add facts which may
help the backend. This is specially useful to interpret the features
of the richer logic implemented by the interactive prover, into the
weaker fragment understood by the automated prover. The second one
works by \emph{translating the goal} $G$ into a possibly different
$G'$, so as to align the definitions used in the goal with those used
by the backend. This essentially involves casting types and signatures
into equivalent ones.

The rest of the section presents a suite of such proof-producing,
predictable, small-scale goal transformations, respectively concerned
with inductive types, with first-order logic and with symbol
interpretation. These transformations are summed-up in {\bf
  Table~\ref{table:transfos}}. They are clearly not exhaustive, but
represent an essential basis of transformations to handle Coq goals,
since they deal with everyday constructions. Future work includes
extending this suite to handle more and more goals; the compositionality
of the approach eases this extension.

The last subsection~(\ref{sec:preproc:metaprogramming}) details the
meta-programming tools that we have used to develop these
transformations.

\begin{table*}[t]
  \begin{center}
  \begin{tabular}{|l|l|l|}
    \hline
    Category & Specification & Tactic name (when relevant for the article) \\
    \hline
    \hline
    Inductive types (\ref{sec:preproc:transfos:indu}) & Inversion principle & \lstinline!inv_principle_all! \\
     & Algebraic data types & - \\
     & Generation statement & \lstinline!get_gen_statement_for_variables_in_context! \\
     & Pattern matching & - \\
    \hline
    Going first-order (\ref{sec:preproc:transfos:fol}) & Equalities & - \\
     & Monomorphization & - \\
    \hline
    Symbols (\ref{sec:preproc:transfos:symbols}) & Constants and fixpoints (\ref{par:preproc:transfos:def}) & - \\
    & Types and logical connectives (\ref{sec:preproc:transfos:trakt}) & \lstinline!trakt! $\bullet$ $\bullet$ \\
    \hline
  \end{tabular}
  \caption{List of transformations}
  \label{table:transfos}
  \end{center}
\end{table*}

\subsection{Axiomatizing inductive types}
\label{sec:preproc:transfos:indu}

All the transformations of this subsection add statements in the proof context. 
As they should be applied to several terms, they come in two versions: 

\begin{enumerate}
\item the \emph{elementary} one taking a term as parameter and applying the transformation to it;
\item the one which scans the proof context and calls 1. on all terms of the suitable form. 
We call it the \emph{context-handling} tactic, but we should emphasize the fact that it acts only 
on the local context and not on the global one. 
\end{enumerate}

\paragraph{Inversion principle of inductive relations}
\label{sec:preproc:transfos:indu:invprinc}

An inductive relation is a \Coq inductive type whose codomain is in 
\Prop. 
Let us consider an inductive relation \lstinline!R! with arguments 
\lstinline!x1 ... xn!. 
In user-written \Coq proofs, the tactic \lstinline!inversion H! is used to 
retrieve how a hypothesis \lstinline!H! of type \lstinline!R x1 ... xn! could 
have been obtained. This is a consequence of the semantics of the inductive type
\lstinline!R!. But some automation tactics will not perform  
\lstinline!inversion! or will not find how to use inversion lemmas 
properly. 
In addition, intuitionistic external tools may need information 
about the symbol \lstinline!R! in order to use it effectively. 
This is the reason why we implemented a tactic which generates and 
proves these inversion lemmas. 
As an example, let us consider the simple relation \lstinline!add!, 
which holds between three natural numbers when the third argument is 
the sum of the first one and the second one:
\begin{lstlisting}
Inductive |*add*| : nat -> nat -> nat -> Prop :=
  | add0 : forall n, add 0 n n
  | addS : forall n m k, add n m k -> add (S n) m (S k).
\end{lstlisting}
After calling the elementary tactic on \lstinline!add!, this new statement is 
added to the local context: 
\begin{lstlisting}
forall (n m k : nat), add n m k <->
  (exists (n' : nat), n = 0 /\ m = n' /\ k = n') \/
  (exists (n' m' k' : nat),
    add n' m' k' /\ n = S n' /\ m = m' /\ k = S k')
\end{lstlisting}
The context-handling tactic generates this principle for all inductive relations encountered in 
the local context and in the goal, and it is called \lstinline!inv_principle_all!.
It is used in example~\ref{ex:exinv1}.

\paragraph{Interpretation of algebraic data types}

Algebraic data types are inductive types with possibly prenex polymorphism 
(parameters) and no other type dependency (\eg indices).
This transformation takes an inductive type $T$, generates 
three families of statements and proves them:
\begin{itemize}
\item $D_{T}$: 
the direct images of the constructors of $T$ are \textit{pairwise disjoint} (no-confusion property);
\item $I_{T}$: the constructors of $T$ are \textit{injective};
\item $G_{T}$: each term of type $T$ is \textit{generated} by one of 
the constructors (generation principle).
\end{itemize}
These three statements axiomatize $T$ in a first-order logic with prenex polymorphism.

\begin{example}
\label{ex:datatypes:list}
In order to illustrate the transformation, we consider the case of the \lstinline!list! inductive type
and we obtain\footnote{$G_\texttt{list}$ can be treated slighly 
differently, see in \ref{para:gen}.}:

\begin{lstlisting}
D_list : forall A (l : list A) (x : A), [] <> x :: l.
I_list : forall A (l l' : list A) (x x' : A),
  x :: l = x' :: l' -> x = x' /\ l = l'.
G_list : forall A (l : list A),
  l = [] \/ exists (x : A) (l' : list A), l = x :: l'.
\end{lstlisting}

Without these statements, many external backends would consider \lstinline!list! 
as an uninterpreted data type and, for instance, would not be able to perform case 
analysis on a list.
\end{example}

\paragraph{Generation statement for inductive types}
\label{para:gen}

\newcommand{\proj}{\texttt{proj}\xspace}



This transformation provides an alternative generation statement for inductive types, 
avoiding existential quantifiers, as it can be an obstacle to some
automated backends.
%
To achieve this, the idea is to introduce new definitions in the local context for \emph{projections.} 
In general, the transformation introduces one such projection
per argument of each constructor of a given inductive type. Then, 
projections are used to describe how the terms of the inductive type can be generated from its constructors.
This is best illustrated  with an example, such as (once again) the 
type \lstinline!list!.
When given a type \code{A},
the transformation \lstinline!get_projs! generates and proves this statement:

\begin{lstlisting}
forall (l : list A) (a : A),
  l = [] \/ l = proj21 A a l :: proj22 A [] l
\end{lstlisting}

Suppose that our inductive type features $n$ constructors
$C_1,\ldots,C_n$ and each constructor $C_i$ has $k_i$ arguments
(excluding parameters). The function for the $i$-th constructor, and the
$j$-th argument (with $j < k_i$), performs a pattern matching on the
inductive term. If this term corresponds to the $i$-th constructor, the
function $\proj_{i,j}$ returns its $j$-th projection, \ie
$\proj_{i,j}\ (C_i\ \ldots\ x_j\ \ldots) = x_j$ .
Otherwise, it returns a default term of the expected type, \ie
$\proj_{i,j}\ (C_{i'}\ \ldots)=\texttt{default}$. This default term may
be found automatically by an auxiliary tactic, or left as a subgoal to
the user.

Thus, $k_1+\ldots + k_n$ projections are generated.
In the case of \lstinline!list!, we have:
\begin{lstlisting}
proj21 := fun (A : Type) (default : A) (l : list A) =>
  match l with
  | [] => default
  | x :: xs => x
  end
proj22 := fun (A : Type) (default : list A) (l : list A) =>
  match l with
  | [] => default
  | x :: xs => xs
  end
\end{lstlisting}

These projections are uninterpreted symbols for the external solver, but
they are defined in \Coq in order to prove the correctness of the
application of the transformation. An automated backend can perform the
case disjunction over a term once the statement (such as the one above)
has been generated and \emph{without} knowing the definitions of the
projections.

To sum up, the elementary transformation works in three steps:
\begin{enumerate}[topsep=0pt]
\item It generates the projections and poses them in the local context;
\item It generates and proves a first generation statement which abstracts over the default terms;
\item Whenever it is possible, it finds an inhabitant for the considered inductive type and generates a statement with no quantification over default terms.
\end{enumerate}

The context-handling one comes in two versions~: either it produces and proves 
the generation statement for all algebraic datatypes \lstinline!I! such that 
there is a statement \lstinline!x : I! in the local context with \lstinline!x! a variable, 
or it specializes the statement for each such variables, 
generating as many corresponding instances. 
In this case, we eliminate the first universal quantifier 
and this can help some automated backends, as in example~\ref{ex:exgen1}, and the tactic is called
\lstinline!get_gen_statements_for_variables_in_context!.

\paragraph{Elimination of pattern matching}

Whenever a hypothesis contains pattern matching, this transformation 
states and proves one statement for each pattern. 
For instance, if we have a hypothesis stating 
the definition of the \lstinline!nth_default! function from \Coq's standard library:
\begin{lstlisting}
forall A d l n, @nth_default A d l n =
  match nth_error l n with
  | Some x => x
  | None => d
  end
\end{lstlisting}
it is replaced with the following two hypotheses: 
\begin{lstlisting}
forall A d l n,
  nth_error l n = Some x -> nth_default d l n = x.
forall A d l n,
  nth_error l n = None -> nth_default d l n = d.
\end{lstlisting}

\subsection{Going first order}
\label{sec:preproc:transfos:fol}

Some tactics of our pre-processing tool are designed to transform or
eliminate constructions or features of CIC that an automated solver may
not be able to interpret. As we want to be small-scale, such
transformations should deal with one particular aspect of \Coq logic at
a time. We currently have two transformations, to deal with higher-order
equalities and prenex polymorphism.

\paragraph{Higher-order equalities}

Whenever a hypothesis in the local context of the proof is of the form
\lstinline!f = g!, this transformation replaces it by quantifying over
the arguments:

\begin{lstlisting}
forall x1 ... xn,  f x1 ... xn = g x1 ... xn
\end{lstlisting}

This is particularly helpful to clarify a hypothesis which recovers the
definition of a function (as will be presented
in~\Sec~\ref{par:preproc:transfos:def}). For instance, from an
hypothesis which simply expands the definition of the
\lstinline!hd_default! function:
\begin{lstlisting}
hd_default = fun (A : Type) (l :list A) (default : A) =>
  match l with
  | [] => default
  | x :: _ => x
  end
\end{lstlisting}
the tactic generates and proves:
\begin{lstlisting}
forall (A : Type) (l : list A) (default : A),
  @hd_default A l default  =
    match l with
    | [] => default
    | x :: _ => x
    end
\end{lstlisting}

This transformation weakens the hypothesis in a Coq point of view, but
goals whose proof require the higher-order equality would need further
transformations to be discharged to backend automation engines based on
first-order logic anyway.

\paragraph{Monomorphization}

Most automated provers do not handle polymorphism at all. This is a
problem as \Coq lemmas are often polymorphic and used later on
monomorphic instances. For this reason, we implemented a tactic which
instantiates all the polymorphic hypotheses given to the tactic and
present in the local context with chosen ground types.

The chosen instances depend on each polymorphic hypothesis, using the following heuristic.
Suppose that the hypothesis \lstinline!H! quantifies over $n$ type variables $A_{1},\ \ldots,\ A_{n}$, and \lstinline!H! contains as a subterm an applied inductive 
of the form $I \: B_{I_{1}} \: \ldots \: B_{I_{k}}$
with codomain \lstinline!Type! and with $k$ parameters of type \lstinline!Type!. 
Then for each $B_{I_{i}}$ being among the
$A_{1}, \: \ldots, \: A_{n}$ (say that $B_{I_{i}}=A_{j}$)
we search in the goal or in the other (monomorphic) hypotheses for a subterm of the form 
$I \: t_{I_{1}} \: \ldots \: t_{I_{k}}$ or $C \: t_{I_{1}} \: \ldots \: t_{I_{k}}$
if $C$ is a constructor of $I$. 
The $i$-th argument of $I$ which is $t_{I_{i}}$ will be an instance for the variable $A_{j}$. 
We scan all the monomorphic hypotheses and the goal in the same way to find all the instances.

\begin{example}
\label{ex:instances}
If we consider the following proof context:
\begin{lstlisting}
H : forall (A B : Type) (x1 x2 : A) (y1 y2 : B),
  (x1, y1) = (x2, y2) -> x1 = x2 /\ y1 = y2
___________________________________________________
forall (x1 x2 : option Z) (y1 y2 : list unit),
  (x1, y1) = (x2, y2) -> x1 = x2 /\ y1 = y2
\end{lstlisting}
the hypothesis contains one type that takes polymorphic parameters: 
\lstinline!*! (the non dependent product type). 
Thus, our tactic should look at how the parameters of \lstinline!*! are 
instantiated in the other hypotheses or in the goal. 
Here, there is no other hypotheses so we look at the goal. 
\lstinline!A! can only be instantiated by \lstinline!option Z! and \lstinline!B! 
by \lstinline!list unit!.
\end{example}

We also use the version of the tactic implemented in
\cite{DBLP:journals/corr/abs-2107-02353}, which is more exhaustive and
faster when the context contains a reasonable number of potential
instances (two or three) but can be slower when the instances are
numerous (exponential in the number of instances).
Bobot and Paskevich proposed a complete transformation to encompass
polymorphism~\cite{DBLP:conf/frocos/BobotP11}. We may also implement it
in next versions, but found in practice our implementation of
monomorphization to be efficient enough, and it has the advantage of
avoiding obfuscating the goal.

\subsection{Giving meaning to symbols}
\label{sec:preproc:transfos:symbols}

The very rich type system of \Coq allows representing theory-specific
values, operations, and predicates with a wide range of data structures.
On the contrary, theory-based automated provers often associate the
signatures of the theories they can process with a few data structures
along with the related values and operations (called \emph{symbols}),
restricting the set of goals that can be mapped without loss to input
statements for the automated provers. In this subsection, we introduce
two pre-processing transformations aiming to bridge this gap, \ie make
automation tactics available to \Coq users regardless of the
representation they use for theory-specific data in their proofs.

\subsubsection{Definitions of functions and constants}
\label{par:preproc:transfos:def}

\paragraph{Expanding constants}

This simple transformation scans the local context and the goal and adds
the definitions of all the functions or constants it encounters. Indeed,
one needs to let automated theorem provers have access to the meaning of
\Coq symbols in order to reason about it. For instance, if the term
considered is the ternary relation \lstinline!in_int!, defined in \Coq's
standard library \lstinline!Arith! and stating that the third integer is
the interval defined by the first one (included) and the second one
(excluded), this tactic adds to the local context a lemma expanding the
definition as an equality:
\begin{lstlisting}
      in_int = fun p q r => p <= r /\ r < q.
\end{lstlisting}

This tactic is modular in the sense that it takes as parameters 
the symbols we do not want to interpret: for instance, 
we do not want the inductive definition of addition to be unfolded in the above fashion,
because most of the backends (\eg SMT solvers) know about arithmetic.

\paragraph{Anonymous fixpoints}
The previous transformation can introduce an anonymous fixpoint when the
constant is recursive. Let us consider the \lstinline!length! function.
The transformation states and proves:
\begin{lstlisting}
length = (fix length_anon := fun (A : Type) (l : list A) =>
  match l with
  | [] => 0
  | _ :: l' => S (length_anon l')
  end)
\end{lstlisting}

For this reason, another transformation replaces the anonymous function
\lstinline!length_anon! with its definition (\lstinline!length!). Once
the higher-order equality is eliminated thanks to the transformation
in~\Sec~\ref{sec:preproc:transfos:fol}, the hypothesis is finally
transformed into:
\begin{lstlisting}
forall (A : Type) (l : list A), length l =
  match l with
  | [] => 0
  | _ :: l' => S (length l')
  end
\end{lstlisting}

\subsubsection{Exploiting the notion of equivalence}
\label{sec:preproc:transfos:trakt}

The next transformation related to symbols is \Trakt, a general goal-rewriting tool powered by user declarations.
The plugin exports commands for the user to declare \emph{translation tuples} (\ie proved associations between source and target symbols), as well as a tactic harnessing these translation tuples during a traversal of the current goal to rewrite it into a pre-processed goal.
The modifications applied to the goal are casting theory-specific values into a designated target type for the theory, and translating logic (connectives, predicates, and relations) from \Prop to \bool or the other way around.
The tactic is certifying, therefore it does not leave proof obligations, and an automation tool can be called right after pre-processing.

In this presentation of \Trakt, we shall illustrate the tool used in association with an SMT-like prover with support for arithmetic, logic, and uninterpreted symbols.
Indeed, this combination of theories being often found in day-to-day \Coq proofs, it is a realistic example.

\paragraph{Equivalent types}

To translate values in a goal from a type $T$ to a type $T'$, the user needs to declare both types as \textit{equivalent types} in \Trakt.
The required declaration for a type equivalence is a translation tuple $(T, T', e, \bar{e}, \text{id}_1, \text{id}_2)$, where $T$ is a source type, $T'$ is a target type, $e : T \rightarrow T'$ and $\bar{e} : T' \rightarrow T$ are embedding functions (\ie explicit casts) in both ways, and $\text{id}_1$ and $\text{id}_2$ are proofs that both of their compositions are identities.

Thanks to this declaration, we can embed every value living in $T$, or any simple functional type containing $T$ (\eg $T \rightarrow$ \bool), into the corresponding target type (\eg $T$ into $T'$, and $T \rightarrow$ \bool into $T' \rightarrow$ \bool).
In particular, \Trakt is able to process uninterpreted symbols and universally quantified variables in a type being the source in an equivalence declaration.

\begin{example}\label{ex:trakt1}

Consider the following goal:

\begin{lstlisting}
forall (P : int -> Prop) (x : int), P x <-> P x
\end{lstlisting}

It contains an uninterpreted predicate \code{P} and two quantifiers.
After the declaration of an equivalence between \code{int} to \code{Z}, \Trakt can pre-process this goal into the one below:

\begin{lstlisting}
forall (P' : Z -> Prop) (x' : Z), P' x' <-> P' x'
\end{lstlisting}

\end{example}

To complete proofs based on a theory, we also need to recognize the \textit{signature} of this theory.
That is, we must be able to translate various operators and values into the ones that the proof automation tool run by the user after \Trakt is able to process, so that they do not remain uninterpreted, while remaining general and independent from the theory.
These operators and values can be declared as \textit{symbols} with a translation tuple $(s, s', p)$, where $s$ and $s'$ are two symbols, and $p$ is an embedding property:
$$
\inferrule
  { s : T_1 \rightarrow \cdots \rightarrow T_n \rightarrow T_o \\\\
    s' : T_1' \rightarrow \cdots \rightarrow T_n' \rightarrow T_o' }
  { p : \forall (t_1 : T_1)\ \cdots\ (t_n : T_n), \\\\
    \HL{e_o^?}\ (s\ t_1\ \cdots\ t_n) = s'\ (\HL{e_1^?}\ t_1)\ \cdots\ (\HL{e_n^?}\ t_n) }
$$

Here, $\HL{e_i^?}$ denotes an optional embedding function from $T_i$ to $T_i'$ (provided that the user declared the equivalence before declaring the symbol), present only if $T_i' \neq T_i$.
This embedding property allows replacing the symbols with their counterparts, introducing or moving embedding functions to preserve typing.

\begin{example}\label{ex:trakt2}

Consider the following goal:

\begin{lstlisting}
forall (f : int -> int -> int) (x y : int),
  f x (y + 0) = f (x + 0) y
\end{lstlisting}

It features universal quantifiers and an uninterpreted function, as in Example~\ref{ex:trakt1}, but also addition and zero on the \code{int} type.
Both symbols can be mapped to their counterparts in \code{Z} after proving their embedding properties\footnote{The scope \code{Z_scope} allows using on \code{Z} the same notations for addition, zero, etc., as for \code{nat}. We disambiguate by expliciting the scope with a postfix \code{\%Z}.}:

\begin{lstlisting}
Lemma |*add_embedding_property*| : forall (x y : int),
  Z_of_int (x + y) = (Z_of_int x + Z_of_int y)%Z.
Lemma |*zero_embedding_property*| : Z_of_int 0 = 0%Z.
\end{lstlisting}

Once all the declarations are made, \Trakt can pre-process the goal into the one below:

\begin{lstlisting}
forall (f' : Z -> Z -> Z) (x' y' : Z),
  f' x' (y' + 0)%Z = f' (x' + 0)%Z y'
\end{lstlisting}

\end{example}

\paragraph{Logic}

In an SMT-like goal, theory-based subterms are logical atoms linked together with connectors, equalities, and various other $n$-ary predicates.
In order to fully process these goals, in addition to theory-based pre-processing, \Trakt is also able to translate these logical values into their Boolean equivalents, and \textit{vice versa}.
They can be declared as \textit{relations} by providing a translation tuple $(R, R', p)$, where $R$ and $R'$ are the two associated relations, and $p$ is a proof of equivalence:
$$
\inferrule
  { R : T_1 \rightarrow \cdots \rightarrow T_n \rightarrow L \\\\
    R' : T_1' \rightarrow \cdots \rightarrow T_n' \rightarrow L' }
  { p : \forall (t_1 : T_1)\ \cdots\ (t_n : T_n), \\\\
    R\ t_1\ \cdots\ t_n\ \HLb{\sim_{L,L'}}\ R'\ (\HL{e_1^?}\ t_1)\ \cdots\ (\HL{e_n^?}\ t_n) }
$$

\begin{center}
\begin{tabular}{|c|c|c|}
\hline
$L$ & $L'$ & $\HLb{\sim_{L,L'}}$\\
\hline
\bool & \bool & $\lambda b.\ \lambda b'.\ b = b'$ \\
\Prop & \bool & $\lambda P.\ \lambda b.\ P \leftrightarrow b = \mathbf{true}$ \\
\Prop & \Prop & $\lambda P.\ \lambda Q.\ P \leftrightarrow Q$ \\
\hline
\end{tabular}
\end{center}

Here, $L$ and $L'$ are logical types (\ie either \Prop or \bool) and $\HLb{\sim_{L,L'}}$ is a way to express equivalence depending on these logical types.
$\HL{e_i^?}$ denotes optional embedding functions, as used above for symbols.

\begin{example}\label{ex:trakt3}

Consider the following goal:

\begin{lstlisting}
forall (f : int -> int) (x : int), f x + 0 = f x
\end{lstlisting}

In addition to the features shown in Examples \ref{ex:trakt1} and \ref{ex:trakt2}, this one features equality on \code{int}, that we might want to turn into a Boolean equality on \code{Z}.
The proof of equivalence in that case is the following:

\begin{lstlisting}
Lemma |*eq_int_equivalence_property*| : forall (x y : int),
  x = y <-> (Z_of_int x =? Z_of_int y)%Z = true.
\end{lstlisting}

Once the declaration is made, \Trakt can pre-process the goal into the one below:

\begin{lstlisting}
forall (f' : Z -> Z) (x' : Z),
  (f' x' + 0 =? f' x')%Z = true
\end{lstlisting}

\end{example}


\paragraph{Beyond type equivalence}

\Trakt also allows declaring \textit{partial} embeddings (\ie non-surjective embeddings), where $e \circ \bar{e}$ is not always an identity.
The declaration for type equivalences can be enriched with an alternative tuple $(T, T', e, \bar{e}, C, p_C, \text{id}_1, \text{id}_{2C})$, where $C$ is a restricting predicate on the embedded values (which we call an \textit{embedding condition}), $p_C$ is a proof that it is true on every embedded value, and $\text{id}_{2C}$ is $\text{id}_2$ restricted to the condition $C$.
When an embedding is partial, every embedding function inserted also adds a condition to the output formula, thus modifying the structure of the goal.

\begin{example}\label{ex:trakt4}

Consider the goal of Example~\ref{ex:trakt3}, with \code{int} replaced with \code{nat}:

\begin{lstlisting}
forall (f : nat -> nat) (n : nat), f n + 0 = f n
\end{lstlisting}

Here are the lemmas the user needs to prove to declare an embedding from \code{nat} to \code{Z}:

\begin{lstlisting}
Lemma |*pC_nat*| : forall (n : nat), (0 <= Z.of_nat n)%Z.
Lemma |*id2C_nat*| : forall (z : Z),
  (0 <= z)%Z -> Z.of_nat (Z.to_nat z) = z.
\end{lstlisting}

Once all the declarations are made, \Trakt can be run to pre-process the goal into the one below (assuming the logical target is \Prop):

\begin{lstlisting}
forall (f' : Z -> Z),
  ((forall (x : Z), 0 <= x -> 0 <= f x) ->
    forall (n' : Z), 0 <= n' -> f' n' + 0 = f' n')%Z
\end{lstlisting}

As we can see, hypotheses were added right after the quantifiers for \code{f'} and \code{n'}, to restrict their domain so that they match the original quantifiers in \code{nat}.

\end{example}

\paragraph{Knowledge database and user API}

Before making proofs, the user can communicate information to \Trakt through four \Coq commands, one for each kind of information declared: integer types, relations, symbols, and terms that can trigger conversion.
Each of these commands is associated to a \coqelpi database with a predicate, and every call to the command adds an instance of the associated predicate to the database.
This allows the user to statically fill the database and then freely call the \trakt tactic that will harness this added knowledge by performing lookups at runtime.

Let us show a simplified syntax of the four available commands.
Integer types are declared through one of the following commands:

\begin{lstlisting}
Trakt Add Embedding T Z e e' id1 id2.
Trakt Add Embedding T Z e e' id1 id2C pC.
\end{lstlisting}

The variable names copy the ones above, except for \lstinline!e'! to replace $\bar{e}$.
Notice that the embedding condition is missing.
It is due to the fact that this type can actually be inferred from $p_C$ or $\text{id}_{2C}$.
This kind of simplification is performed everytime it is possible, to relieve the user from useless repetitions.

For instance, using the lemmas stated in Example~\ref{ex:trakt4}, we can declare an embedding from \code{nat} to \code{Z} with this command:

\begin{lstlisting}
Trakt Add Embedding
  nat Z Z.of_nat Z.to_nat id1_nat id2C_nat pC_nat.
\end{lstlisting}

Relations and symbols follow a similar model (relations need the arity
to be specified because inference is more complex). The user may also
add terms that will explicitely trigger conversion, in order to avoid
being blocked by redefinitions of constant.

\begin{lstlisting}
Trakt Add Symbol s s' p.
Trakt Add Relation n R R' p.
Trakt Add Conversion t.
\end{lstlisting}

\subsection{Meta-programming}
\label{sec:preproc:metaprogramming}

\Coq offers various approaches to \emph{meta-programming}, that is, to implement programs operating on the syntax of arbitrary \Coq terms.
In addition to \Ltac~\cite{the_coq_development_team_2022_5846982}, the default tactic language available in \Coq, used to glue the various transformations together, the present contribution combines two such meta-languages, which offer in particular different levels of control on the syntax for terms: \MetaCoq and \coqelpi.

\subsubsection{MetaCoq}

\MetaCoq~\cite{DBLP:journals/jar/SozeauABCFKMTW20} is a plugin which provides a quoted syntax of \Coq terms, defined as a \Coq inductive type.
This plugin also provides a tactic \lstinline!quote_term!, which turns a Gallina term into its quoted counterpart, and an analogous anti-quotation tactic.
\MetaCoq is useful for performing a very fine-grained analysis on the
syntax of \Coq terms and provides information about the global environment:
we use it to create new statements from a \Coq term (such as in~\ref{sec:preproc:transfos:indu}).
The technicalities of de Bruijn indices and the lack of pretty-printing or notation tools sometimes limit its usability.


\subsubsection{Coq-Elpi}

To build a tool able to perform the \Trakt transformation presented in \ref{sec:preproc:transfos:trakt}, two technical challenges arise: term traversal (\ie recursively inspecting a \Coq term and building a new term at the same time) and user knowledge management (\ie a way to declare translation tuples, a database to store them, and a mechanism to perform lookups during pre-processing).
For the implementation, we used \coqelpi~\cite{tassi:LIPIcs:2019:11084}, an implementation of \lambdaProlog (called \elpi) coupled with an API that connects it to \Coq internals, allowing to make new declarations, commands, and tactics.
Let us now give a few implementation details about the plugin and explain how this meta-language meets our needs.

\paragraph{Term traversal and reconstruction}

In \Trakt, the pre-processing algorithm takes the shape of a \coqelpi predicate with various cases according to the shape of the inspected term.
These cases are based on an inductive type in the meta-language representing \Coq terms in \textit{Higher-Order Abstract Syntax}~\cite{DBLP:conf/cl/Miller00} (HOAS), meaning that terms under binders are represented with meta-functions.

For instance, universal quantifiers are represented with \elpicode{prod N T F}, where \code{N} is the name of the bound variable, \code{T} is its type, and \code{F} is a function (\ie of type \elpicode{term -> term} at the meta level) encoding the quantified term.
To traverse this term, we introduce a locally bound variable \code{x} (called a \emph{universal constant}) thanks to the \elpicode{pi} keyword, then we can recursively traverse \elpicode{F x}.
In particular, this way of binding \Coq variables directly frees us from having to handle de Bruijn indices in the meta-program.
In addition, when using a universal constant \code{x}, we express the outputs of the subsequent calls as functions of \code{x}, ensuring by design that no output term will make the variable escape its scope.
Finally, thanks to \Prolog-like implication, it is possible to assume information about universal constants, adding this information to the scope of subsequent calls, thus eliminating the need to maintain any context in the pre-processing predicate.

To show all these features in a real use case, let us give a simplified snippet taken from the code of \Trakt.
Consider a predicate \lstinline!preprocess! which takes the input goal and outputs the pre-processed goal along with the certification of the goal substitution.
The base cases of the predicate should correspond to transformation tuples, but here is the case for universal quantifiers:

\begin{lstlisting}[language=ELPI]
preprocess (prod N T F) Out Proof :- !,
  pi x\ decl x _ T => preprocess (F x) (F' x) (ProofF x)
  % ...
\end{lstlisting}

As explained above, we introduce a universal constant \code{x} to make a recursive call on \code{F x}, and by using the implication, we introduce \code{decl x _ T} into the scope of the recursive call, thus assuming that the type of \code{x} is \code{T}.
The proof on the subterm is represented by a meta-function \code{ProofF}, so it does not contain \code{x}.
The rest of the code for this case of the predicate focuses on lifting this proof to a proof on the quantified term.

\coqelpi also features a quotation and anti-quotation mechanism, to ease the expression of the output term and the corresponding \Coq proof.
Indeed, it is possible to use \Coq syntax to express a term by surrounding the term with brackets, and it is possible to mention \elpi terms thanks to the \code{lp:} prefix.
For example, \elpicode{\{\{ lp:A -> lp:B \}\}} denotes a \Coq term embedded in \elpi.
It is the type of \Coq functions from a type expressed in the \elpi variable \code{A} to another expressed in the variable \code{B}.

\paragraph{Management of user knowledge}

The \coqelpi meta-language suits perfectly the way the \Trakt plugin
works.  Indeed, it offers a way to declare databases and fill them
with terms extracted from commands typed by the user, and have a
tactic use these databases afterwards.  \Trakt requires the user to
register embeddings, symbols and relations through the available
commands, and the \trakt tactic performs lookups on this database to
pre-process the goal.  To our knowledge, the other common
meta-languages for \Coq do not allow for such a direct implementation
of this information accumulation procedure, and the related user
commands.

\section{From single to composed pre-processing and applications}
\label{sec:applications}

This section revisits the examples discussed in \Sec~\ref{sec:preproc}
and illustrates the benefits brought by the various tactics presented
in \Sec~\ref{sec:preproc}, to various automated backends. These
tactics can be used in isolation, for a single-component
preprocessing, or combined according to a specific
strategy. Section~\ref{sec:preproc:combining} presents an example of
such a combination, designed for the \SMTCoq backend.

\subsection{Single-component pre-processing}
\label{sec:preproc:single}

\subsubsection{Theories
}

The association of the \Trakt tactic with the \itauto
plugin~\cite{DBLP:conf/itp/Besson21} for satisfiability modulo theory,
overcomes the limitations of the latter discussed in
\Sec~\ref{sec:examples}.

\begin{example}\label{ex:trakt}
Here is an example of goal successfully proved by the association of
\Trakt and \itauto, phrased using the \code{int} type from library
\MathComp:

\begin{lstlisting}
Goal forall (f : int -> int) (x : int),
  (f (2%:Z * x) <= f (x + x))%R = true.
Proof. trakt Z Prop. smt. Qed.
\end{lstlisting}
\end{example}

This goal features arithmetic symbols (addition, multiplication), a
Boolean comparison relation, an uninterpreted function \code{f}, and
values in the \code{int} type from \MathComp. Notations are moreover
generic: \code{+} (resp. \code{*}) refers to the law of an instance of
commutative group (resp. the product of a ring) and \code{<=} is a
generic notation for an order relation. Note how the translation is
configured so as to target the \code{Prop} sort, so that the \itauto
solver is eventually able to prove the goal automatically.

Generic notations like the ones of example~\ref{ex:trakt} are
implemented using the form of \emph{ad hoc} polymorphism
available in \Coq via a combination of advanced inference, canonical
structures~\cite{DBLP:conf/itp/MahboubiT13} or
typeclasses~\cite{DBLP:conf/tphol/SozeauO08}, and notations. Dealing
with this overloading requires some care in the
implementation. Indeed, by default, the right case of the
pre-processing predicate is selected by performing a \Prolog-like
unification between the input term and the pattern in the head of the
clause (\eg \elpicode{prod N T F} in the case above). But in this
case, purely syntactic matching does not suffices.

\begin{example}\label{ex:trakt5}

Consider again Example~\ref{ex:trakt3}, where notations \code{+} and
\code{0} are instances of notations based on \emph{ad hoc}
polymorphism (enabled in the \code{ring_scope} notation scope denoted
by the postfix \code{\%R}):

\begin{lstlisting}
forall (f : int -> int) (x : int), (f x + 0)%R = f x
\end{lstlisting}

Term \code{f x + 0} actually unfolds to the term with holes
\begin{lstlisting}
  @GRing.add _ (f x) (@GRing.zero _)
\end{lstlisting}
In this case, holes are filled by \Coq with the canonical instance of
ring structure available for \code{int}, so that in the end
\code{f x + 0} is elaborated as:
\begin{lstlisting}
  @GRing.add int_ZmodType (f x) (@GRing.zero int_ZmodType)
\end{lstlisting}
As a consequence, in this goal, the
addition on type \code{int} is represented as \code{(@GRing.add
  int_ZmodType)}, a term convertible but not syntactically equal to
the defined constant
\begin{lstlisting}
  addz : int -> int -> int
\end{lstlisting} registered as an equivalent to
\begin{lstlisting}
  Z.add : Z -> Z -> Z
\end{lstlisting}

\Trakt supports an additional declaration of terms that can trigger
\Coq conversion, so that the translation algorithm cannot distinguish
the goal above from the one in Example~\ref{ex:trakt3}, thus
successfully pre-processing goals featuring such projections of
instances of structures.  As a result, running \code{trakt Z bool} on this example
gives the expected output goal:

\begin{lstlisting}
forall (f' : Z -> Z) (x' : Z), (f' x' + 0 =? f' x')%Z = true
\end{lstlisting}

\end{example}

\subsubsection{Inversion principle}
\label{sec:preproc:gen:inv}
The first volume of the Software Foundations
textbooks~\cite{Pierce:SF1} uses the following \code{ev} inductive
predicate to illustrate the \emph{inversion} of inductively defined
propositions in the eponymous chapter:

\begin{lstlisting}
Inductive |*ev*| : nat -> Prop :=
  | ev_0 : ev 0
  | ev_SS (n : nat) (H : ev n) : ev (S (S n)).
\end{lstlisting}

\begin{example}\label{ex:exinv1}
A manual proof requires performing inversion twice to prove:
\begin{lstlisting}
Lemma |*SSSSev_ev*| : forall (n : nat),
  ev (S (S (S (S n)))) -> ev n.
Proof.
  Fail firstorder congruence. 
  inv_principle_all; firstorder congruence.
Qed.
\end{lstlisting}
Notwithstanding the pedagogical merits of this exercise in iterating
inversion, one may wish in realistic situations for a script which
does not depend on the number of successor symbols \code{S} involved
in the statement. However, the \firstordercongruence tactic alone is
not able to prove this goal, although the latter falls in the fragment
of equality logic with uninterpreted functions decided by this
tactic. Generating (automatically) the appropriate inversion lemma, about
\lstinline!ev!, is however enough to help the proof-search procedure.
\end{example}

\subsubsection{Generation principle}
\label{sec:preproc:single:gen}
As Blazy and Leroy pointed out when reporting on their work about
memory models for \CompCert~\cite{DBLP:journals/jar/LeroyB08}, most of
the related \Coq developments fall under first-order reasoning and
proofs could be better automated. But the incomplete support for
inductive types of the available automation tactics for first-order
reasoning is often a showstopper. Automating the generation of the
appropriate properties of constructors is however often enough to
enable automation.

\begin{example}\label{ex:exgen1}
In \CompCert, the type \lstinline!memory_chunk! is an enumeration
indicating the type, size and signedness of the chunk of memory being
accessed. It can be translated into an integer by the function
\lstinline!memory_chunk_to_Z!. It associates to the $i$-th constructor
the integer $i$. None of the variants of the \code{sauto} tactic
provided by the \CoqHammer plugin can prove lemma
\code{memory_chunk_to_Z_eq }, but \code{sauto} can prove it after
suitable principles have been generated.

\begin{lstlisting}
Lemma |*memory_chunk_to_Z_eq*| : forall x y,
  x = y <-> memory_chunk_to_Z x = memory_chunk_to_Z y.
Proof.
  Fail sauto. 
  get_gen_statement_for_variables_in_context; sauto.
Qed.
\end{lstlisting}
\end{example}
The same association can prove lemma \code{app_nilI}, as does the
\code{sauto dep: on} variant of \code{sauto}:
\begin{lstlisting}
Lemma |*app_nilI*| : forall B (l1 l2 : list B),
  l1 ++ l2 = [] -> l1 = [] /\ l2 = [].
  Fail sauto.
  get_gen_statement_for_variables_in_context; sauto.
Qed.
\end{lstlisting}

\subsection{Composite pre-processing}
\label{sec:preproc:combining}

\subsubsection{The \snipe tactic}
\label{sec:preproc:combining:sniper}

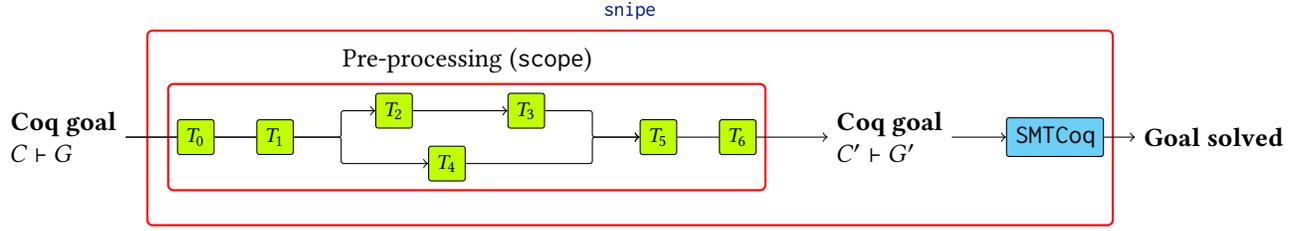
\begin{figure*}[t]
  \begin{center}
    \begin{tikzpicture}[x=5pt,y=-10pt]
      \draw (-10,1) node (CoqGoal) {\vbox{\hbox{\textbf{Coq goal}}\hbox{$C \vdash G$}}};
      \draw[rounded corners=1pt] (0,1) node[draw, fill=lime] (transf0) {\footnotesize $T_0$};
      \draw[rounded corners=1pt] (6,1) node[draw, fill=lime] (transf1) {\footnotesize $T_1$};
      \draw (11,1) node[shape=coordinate] (split) {};
      \draw[rounded corners=1pt] (15,0) node[draw, fill=lime] (transf2) {\footnotesize $T_2$};
      \draw[rounded corners=1pt] (25,0) node[draw, fill=lime] (transf3) {\footnotesize $T_3$};
      \draw[rounded corners=1pt] (19,2) node[draw, fill=lime] (transf4) {\footnotesize $T_4$};
      \draw (30,1) node[shape=coordinate] (merge) {};
      \draw[rounded corners=1pt] (35,1) node[draw, fill=lime] (transf5) {\footnotesize $T_5$};
      \draw[rounded corners=1pt] (41,1) node[draw, fill=lime] (transf6) {\footnotesize $T_6$};
      \draw (52.5,1) node (CoqGoal2) {\vbox{\hbox{\textbf{Coq goal}}\hbox{$C' \vdash G'$}}};

      \draw[rounded corners=1pt] (65,1) node[draw, fill=cyan!50] (SMTCoq) {\SMTCoq};
      \draw (77,1) node (GoalSolved) {\textbf{Goal solved}};
      \draw[thick] node[draw=red,rounded corners=2pt, fit=(transf4) (transf2) (transf6) (transf0), label=Pre-processing (\scope)
      ] {};
      \draw (-3,-2.7) node[shape=coordinate] (adjustOuterBox1) {}; 
      \draw (30,4) node[shape=coordinate] (adjustOuterBox2) {};  
      \draw[thick] node[draw=red,rounded corners=2pt, fit=(transf4) (transf2) (transf6) (transf0) (SMTCoq) (adjustOuterBox1) (adjustOuterBox2), label={\large \snipe}
      ] {};
      \draw[->,rounded corners=1pt] (CoqGoal) -- (transf0) -- (transf1) -- (split) |- (transf2);
      \draw[->] (transf2) -- (transf3);
      \draw[->,rounded corners=1pt] (CoqGoal) -- (transf0) -- (transf1) -- (split) |- (transf4);
      \draw[->,rounded corners=1pt] (transf3) -| (merge) -- (transf5);
      \draw[->,,rounded corners=1pt] (transf4) -| (merge) -- (transf5);
      \draw[->] (transf5) -- (transf6) -- (CoqGoal2);
      \draw[->] (CoqGoal2) -- (SMTCoq);
      \draw[->] (SMTCoq) -- (GoalSolved);

    \end{tikzpicture}
    \caption{The architecture of the \snipe tactic}
    \label{fig:methodology}
  \end{center}
  \vspace*{-0.4cm}
\end{figure*}

The \Sniper plugin~\cite{DBLP:journals/corr/abs-2107-02353} equips the
\SMTCoq backend\footnote{It targets more particularly the \verit tactic,
  which calls the external SMT solver
  \veriT~\cite{DBLP:conf/cade/BoutonODF09}.} with a pre-processing
tactic called \scope. The general methodology of this plugin is
presented in {\bf Figure~\ref{fig:methodology}}: the \scope
pre-processing tactic chains small-scale transformations (represented as
$T_0 \dots T_6$ in the figure), possibly with various paths depending on
the goal and context, then the \SMTCoq automated backend is called.

We have significantly improved \scope, which now incorporates all (but
one, specific to intuitionistic logic) of the transformations presented
in \Sec~\ref{sec:preproc}, and relies on an enhanced strategy for
orchestrating their combination.
The strategy implemented by this new version of \scope proceeds as
follows.  First, \Trakt translates each inductive predicate in
\lstinline!Prop! present in the goal into its Boolean fixpoint
equivalent, if available in its database.  Then, the context is
enriched with first-order polymorphic statements about the functions
and algebraic data types present in the goal, so as to provide \veriT
with the needed part of the \Coq environment. Then, these statements
are instantiated and finally, using \Trakt again allows us to target
the type \lstinline!bool! of classical propositions and the type
\lstinline!Z! of relative integers, so as to meet \SMTCoq's
requirements.  Note that the transformation which generates the
inversion principles of inductive relations is not used by \scope, as
it is rather designed for backends dealing with intuitionistic logic.
This strategy currently freezes an order for the transformations (that
is to say a path in the general methodology of {\bf
  Figure~\ref{fig:methodology}}); we live for future work to improve
flexibility (see Section~\ref{sec:conclusion}).

With this new pre-processing, the \snipe tactic automatically solves all
the goals of the previous subsection but \lstinline!SSSSev_ev!. We now
showcase it on two 
\Coq developments: a formalization of interval lists, and a
formalization of a $\lambda$-calculus based on De Bruijn indices.

\subsubsection{Use case of \snipe: interval lists}
\label{sec:preproc:combining:intlist}

This use case deals with a formalization in \Coq of interval
lists~\cite{LedJFLA2020}, representing constraints on integer
variables. This example is particularly well-suited for SMT-based
automation, as it involves data structures with a decidable theory,
and linear arithmetic. This development represents domains
by a \Coq inductive type:

\begin{lstlisting}
Inductive |*elt_list*| :=
  | Nil : elt_list
  | Cons : Z -> Z -> elt_list -> elt_list.
\end{lstlisting}


where \code{Z} is the integer type from the standard library. These
intervals come with a well-formedness condition, formalized as an
inductive relation, ensuring that the intervals which belong to the
list are in ascending order, disjoint and not empty. The first
argument of the relation, of type \lstinline!Z!, is the lower bound of
the interval list.

\begin{lstlisting}
Inductive |*Inv_elt_list*| : Z -> elt_list -> Prop :=
  | invNil  : forall b, Inv_elt_list b Nil
  | invCons :
    forall (a b j : Z) (q : elt_list),
      j <= a -> a <= b ->  Inv_elt_list (b + 2) q ->
      Inv_elt_list j (Cons a b q).
\end{lstlisting}

It is possible and helpful to write a Boolean equivalent
\lstinline!Inv_elt_list_bool! of this predicate:


\begin{lstlisting}
Lemma Inv_elt_list_decidable : forall b e, 
  Inv_elt_list b e <-> Inv_elt_list_bool b e = true.
\end{lstlisting}
In particular, this Boolean variant is well-suited for the \SMTCoq{}
backend, based on classical logic. At this point, the correspondance
between the two equivalent versions of the well-formedness condition
can be added to \Trakt's database by using the command:
\begin{lstlisting}
Trakt Add Relation
  Inv_elt_list Inv_elt_list_bool Inv_elt_list_decidable.
\end{lstlisting}

Whenever the \trakt tactic targets \code{bool}, any goal featuring
\code{Inv_elt_list} will be translated into a goal featuring its
Boolean counterpart \code{Inv_elt_list_bool}. Then, the
transformations of \scope about interpreted symbols can state and
prove properties about this Boolean counterpart in an understandable
way for a SMT solver, and in particular, for \SMTCoq.

\begin{example}
The following monotonicity property:
\begin{lstlisting}
Lemma |*inv_elt_list_monoton*| : forall l y z,
  Inv_elt_list y l -> z <= y -> Inv_elt_list z l.
Proof.
  induction l; snipe.
Qed.
\end{lstlisting} 
can be proved by an induction on the list of intervals \lstinline!l!,
for which sub-cases are solved by the \snipe tactic.
\end{example}

The same development introduces the domain of a list of interval,
represented as a \Coq record:
\begin{lstlisting}
Record |*t*| := mk_t
  { domain : elt_list;
    size : Z;
    max : Z;
    min : Z; }.
\end{lstlisting}
equipped with the following well-formedness condition:

\begin{lstlisting}
Definition |*Inv_t*| (d : t) := 
  Inv_elt_list (min d) (domain d) /\
  (min d) = get_min (domain d) min_int /\
  (max d) = process_max (domain d) /\
  (size d) = process_size (domain d).
\end{lstlisting}

More explicitly, the domain should verify the previous invariant 
\lstinline!Inv_elt_list!, the 
value corresponding to the field \code{min} should be the same as the one 
computed by the \lstinline!get_min! function, and similar conditions
apply to the fields \lstinline!size! and \lstinline!max!. The functions 
computing the minimum and the maximum of the list of intervals are 
returning a default value called \lstinline!min_int! whenever the 
domain is empty. The constant \lstinline!min_int! is an unconstrained 
parameter of the formalization.
 
Notice that the only correct term of type \lstinline!t! with an empty
domain is the following:

\begin{lstlisting}
Definition |*empty*| :=
  {| domain := Nil; size := 0;
     max := min_int; min := min_int |}.
\end{lstlisting}

After introducing these new definitions, and the Boolean counterpart
of the well-formedness condition, \snipe is able to prove
automatically facts like:

\begin{lstlisting}
Lemma |*empty_inv*| : Inv_t empty. Proof. snipe. Qed.

Lemma |*equiv_empty_Nil*| : forall d : t,
  Inv_t d -> domain d = Nil <-> d = empty.
Proof. snipe. Qed.
\end{lstlisting}

\subsubsection{Use case of \snipe: De Bruijn indices}
\label{sec:preproc:combining:debruijn}

This example deals with a formalization in \Coq of confluence and
strong normalization for some $\lambda$-calculi~\cite{lambda}, based
on the \MathComp libraries. This formalization involves deeply
embeddings of languages with binders, and uses de Bruijn indices to
encode bound variables. This guarantees in particular the uniqueness
of term representation. The price to pay is the need to prove
cumbersome properties on lifting and substitution of variables. Such
proofs typically involve a combination of integer arithmetic and
propositional reasoning, together with induction with a well-chosen
set of generalized variables.

For instance, untyped $\lambda$-calculus is defined as:
\begin{lstlisting}
Inductive |*term*| : Type :=
  | var of nat
  | app of term & term
  | abs of term.
\end{lstlisting}
with the lifting and substitution functions being respectively:
\begin{lstlisting}
Fixpoint |*shift*| d c t : term :=
  match t with
  | var n => var (if c <= n then n + d else n)
  | app t1 t2 => app (shift d c t1) (shift d c t2)
  | abs t1 => abs (shift d c.+1 t1)
  end.
\end{lstlisting}
and
\begin{lstlisting}
Notation substitutev ts m n :=
  (shift n 0 (nth (var (m - n - size ts)) ts (m - n))).
Fixpoint |*substitute*| n ts t : term :=
  match t with
  | var m => if n <= m then substitutev ts m n else m
  | app t1 t2 =>
    app (substitute n ts t1) (substitute n ts t2)
  | abs t' => abs (substitute n.+1 ts t')
  end.
\end{lstlisting}
Observe that these definitions make use of the \MathComp definitions
of addition, subtraction and comparison on natural numbers. By adding
them to the database of \Trakt, an induction on terms followed by the
\snipe tactic is sufficient to prove a number of the aforementioned
properties, such as the following\footnote{We display the original scripts, written using the ssreflect tactic language. Comments provide an analogue version in vanilla Coq.}:
\begin{lstlisting}
Lemma |*shift_add d d'*| c c' t :
c <=? c' -> c' <=? c + d ->
shift d' c' (shift d c t) = shift (d' + d) c t.
(* synonym of: revert d d' c c'; induction t; snipe. *)
Proof. elim: t d d' c c'; snipe. Qed.

Lemma |*shift_shift_distr*| d c d' c' t :
c' <=? c ->
shift d' c' (shift d c t) =
                          shift d (d' + c) (shift d' c' t).
(* synonym of: revert d d' c c'; induction t; snipe. *)
Proof. elim: t d d' c c'; snipe. Qed.
\end{lstlisting}

\section{Related work and perspectives}
\label{sec:conclusion}

The design of user-friendly yet robust automation tools for formal
proofs is inherently difficult a problem. For instance, users might
not realize that a minor change in a formula might put their problem
in an undecidable fragment. As a consequence, some authors like
Chlipala advocate a development model for formal libraries in which
each new project comes with its own tactic library, with ``a different
mix of undecidable theories where a different set of heuristics turns
out to work well''~\cite{DBLP:books/daglib/0035083}. The present paper
proposes a collection of general-purpose, small-scale, and proof-producing goal
transformations, listed in \Sec~\ref{sec:preproc}, which can
contribute in a compositional and reusable way to such
project-specific tactics. In particular, \Sec~\ref{sec:applications}
illustrates how these transformations can be associated with various
automated reasoning backends, for applications in various
contexts. Targeting more heuristic backends would be just as
easy. Because these transformations are small-scale and proof-producing, they do
not blur diagnosis in case of failure of the resulting automatic
tactic. Because they are small scale and compositional, they can be
orchestrated according to different strategies, depending on the
project or on the backend. They are independent from other features of
hammers, like the ability to select relevant facts from the context,
but can greatly help proof reconstruction.

The present work extends and generalizes a previous
work~\cite{DBLP:journals/corr/abs-2107-02353} that combined some
pre-processing tactics with the \SMTCoq automation tool. Notably,
pre-processing is now completely untied from the backend, which allows
for an association with arbitrary tactics, as opposed to just the
\SMTCoq plugin. Moreover, the pool of pre-processing tactics has been
augmented, in particular with the support for translating along
declared equivalences of types or symbols~\cite{trakt}, and for inductive
relations. The support for algebraic data types has also been
improved.

The generation of auxiliary principles for inductive types, for
helping programs and proofs, automated or manual, is a classic
topic~\cite{DBLP:conf/types/McBrideGM04} and different
meta-programming techniques have been studied in particular for the
generation of inductive schemes~\cite{tassi:LIPIcs:2019:11084}.

By contrast, the only work we are aware of about independent,
theory-specific pre-processing is the \lstinline+zify+~\cite{ppsimpl}
tactic.  For instance, the \lstinline+mczify+ library declares the
relevant instances to data structures defined in the \MathComp library
so as to bring the power of the \code{lia} tactic, automating linear
integer arithmetic, to users of these libraries. As illustrated in
\Sec~\ref{sec:examples}, the pre-processing implemented by \code{zify}
is quite specific and cannot in general traverse uninterpreted
symbols. Tools for normalizing coercions and
casts~\cite{DBLP:conf/cade/LewisM20} can also be seen as
pre-processing devices.

Generalizing the latter issue with coercions and casts, different
design patterns in the hierarchies of algebraic structures, underlying
generic notations, may also limit the usage of the associated decision
procedures (e.g., for ring equalities): this problem is discussed and
solved by Sakagushi, using an extra layer of
reflexion~\cite{DBLP:conf/itp/Sakaguchi22}.

Univalent foundations are
meant to better account for the transfer of properties from one
representation of a mathematical concept to another, equivalent one.
Homotopy equivalences, at the core of homotopy type theory, make
precise the types for which any such property can be
transferred. Under the assumption of the univalent axiom, univalent
parametricity makes possible to perform this transfer in practice, as
partially implemented in the companion prototype to Tanter et al.'s
paper~\cite{10.1145/3429979}. Although illustrating nicely the paper's
contributions, the latter prototype is however not usable beyond toy examples.
However, a consolidated re-implementation of the same paradigm could
provide a more principled implementation of the \code{trakt} plugin
for proofs under the univalence assumption.

Formal proof automation based on the cooperation with a first-order
automated reasoning device usually encompasses some notion of
pre-processing, although as an indivisible whole. This pre-processing
part is then usually tied to the backend(s) used, as, e.g., in the
case of \CoqHammer~\cite{DBLP:journals/jar/CzajkaK18} or of the
\fstar~\cite{DBLP:conf/popl/SwamyHKRDFBFSKZ16} programming language.
In the case of \fstar however, a tactic language was added to help
users target (solely) the SMT
backend~\cite{DBLP:conf/esop/MartinezADGHHNP19}.  In comparison, we
use here the \Coq tactic language and meta-programming tools, and we
target various backends.

Backends are willing to evolve. For instance, the efficiency of SMT solvers such as
\veriT and \cvcfive on --certain subsets of-- higher-order
logic is gaining traction~\cite{DBLP:conf/cade/BarbosaROTB19}. Our pre-processing approach
is well-suited to follow such developments, as small-scale
transformations can be easily plugged or unplugged.

Expanding further the pool of small-scale transformations, and their
efficiency, is a natural direction of future work. Another direction,
may be even more important, is to study more in-depth the appropriate
strategies for organizing of the cooperation of small-scale
tactics. Ideally, less expertise in meta-programming should be
required for implementing the latter strategies. Designing a suitable
API would help creating new pre-processing tactics, either by a
fine-grained description of the actions, or by triggering automated
patterns from an inspection of the goals, or by a combination thereof.

\begin{acks}
  The authors would like to thank Kazuhiko Sakaguchi, for his help on
  the example of Section~\ref{sec:preproc:combining:debruijn} and his
  proofreading of the paper, and Amélie Ledein, for her help on the
  example of Section~\ref{sec:preproc:combining:intlist}.

  This work was partially funded by a Nomadic Labs-Inria collaboration.
\end{acks}

\bibliographystyle{ACM-Reference-Format}
\bibliography{biblio}

\end{document}